\documentclass[11pt,twoside]{article}
\usepackage{asp2010}

\resetcounters

\begin{document}

\title{Narrow Polar Rings versus Wide Polar Ring/Diks Galaxies}

\author{E.~Iodice}
\affil{INAF-Osservatorio Astronomico di Capodimonte, Napoli, Italy}

\begin{abstract}
In the latest ten years, a big effort has given to study the
morphology and kinematics of polar ring galaxies: many steps forward
and new discoveries on the structure and formation mechanisms for such
systems have been made during this time thanks to high resolution
photometric and spectroscopic data. In this paper, I review the latest
results obtained for this class of galaxies, from both observational
and theoretical studies. I focus on the analysis of the observed
properties (e.g., structure, colours, age, metallicity, and
kinematics) for narrow and wide polar ring galaxies. In particular, I
compare AM~2020-504 and NGC~4650A, which are the two prototypes for
narrow and wide polar rings, respectively. I discuss similarities and
differences between the two kinds of systems and how they reconcile
with the main formation scenarios proposed for this class of galaxies.
\end{abstract}

\section{Introduction}

Polar ring galaxies (PRGs) are multi-spin systems composed by a
central spheroidal component, the host galaxy (HG), and a polar
structure which orbits in a nearly perpendicular plane to the
equatorial one of the HG. In 1990, Whitmore and collaborators compiled
the Catalog of Polar Ring Galaxies, Candidates and Related Objects
(PRC), which includes 157 galaxies, where 106 have a well-defined PRG
morphology, i.e., the two components (HG and polar structure) are
clearly detectable and/or kinematically decoupled. In the rest of the
sample, galaxies show a perturbed structure, where the central object
has polar or high-inclined feature. Recently, taking advantage of new
large data set from surveys, the number of PRG candidates PRG grows
up, also at higher redshifts than galaxies in the PRC \citep[][see
also Iodice et al., this volume]{Finkelman,Resh1997a}. In
particular, \citet{Moiseev2011} made a new catalogue of PRGs based on
the Sloan Digital Sky Survey (SDSS), providing new 275 candidates for
polar rings and related objects (see also Smirnova et al., this
volume).

For all the included objects in the PRC, the polar structure is
identified as a {\it ring\/} \citet{Whitmore1990}. Only subsequent
studies on the prototype of PRGs, NGC~4650A (Fig. \ref{fig1}), have
revealed for the first time that the polar structure in this object
has the morphology and kinematics of a {\it disk\/}, rather than a ring
\citep[see][]{Arnaboldi1997, Iodice02, Gallagher2002, Swaters2003}. 
NGC~4650A does not seem the only polar disk galaxy in the universe,
but other PRGs show similar characteristics, such as A0136-0801 (PRC
A-1) and UGC~9796 (PRC A-6) in PRC and SPRC-27, SPRC-59, and SPRC-69
in the new SDSS-based Polar Ring Catalog \citep[SPRC;][]{Moiseev2011}.
Thus, the `PRG' morphological type currently includes both polar rings
and polar disks. As I show in this review, they have a different
structure (i.e., light distribution, colors, age, and star motions)
and probably a different formation history (but this is still an open
issue).

{\it Why it is interesting to study PRGs?\/} Given the unique
geometry, as suggested by \citet{Whitmore1990}, PRGs represent a new
way to study galaxy structure and formation. The existence of two
orthogonal components of the angular momentum makes the PRGs the ideal
laboratory to derive the 3-dimensional shape of the potential (see
Combes, this volume).  The multi-spin morphology can not be explained
by the collapse of a single proto-galactic cloud, but some kind of
interaction (galaxy-galaxy or galaxy-environment) need to be invoked
in the formation history of these systems.  The gravitational
interactions are the `carrying pillar' of the cold dark matter (CDM)
model for galaxy formation \citep[e.g.,][]{Col00}, and in this
framework, both merger and gas accretion play a major role in building
the structure of spheroid and disk (see Conselice, this volume).
Thus, PRGs are among the best objects in the universe to study the
physics of such processes.  In PRGs, as well as for all types of
galaxies, the observed morphology, kinematics, structure, and physical
properties (i.e., stellar population, gas and dust content, and
metallicity) provide the record of the formation mechanism. They
result from the relative contribution of each kind of interaction
(i.e., merger, accretion, and tidal stripping), from the physical
parameters (i.e., mass and gas ratios) and orbital configuration
(i.e., trajectory and relative velocity) of the two (or more)
interacting systems.  All the observed physical quantities listed
above need to be reproduced by models of galaxy formation.

In this paper, I review the main observational properties of galaxies
with narrow and wide polar rings/disks (Sect.~\ref{obs}), which are
the most reliable formation scenarios proposed for PRGs
(Sect.~\ref{theory}) and how well they reconcile with the observations
(Sect.~\ref{conf}).

\section{Narrow versus Wide Polar Rings/Disks: Observational Properties}
\label{obs}

\begin{figure*}[!t]
\centering
\includegraphics[width=13cm,keepaspectratio]{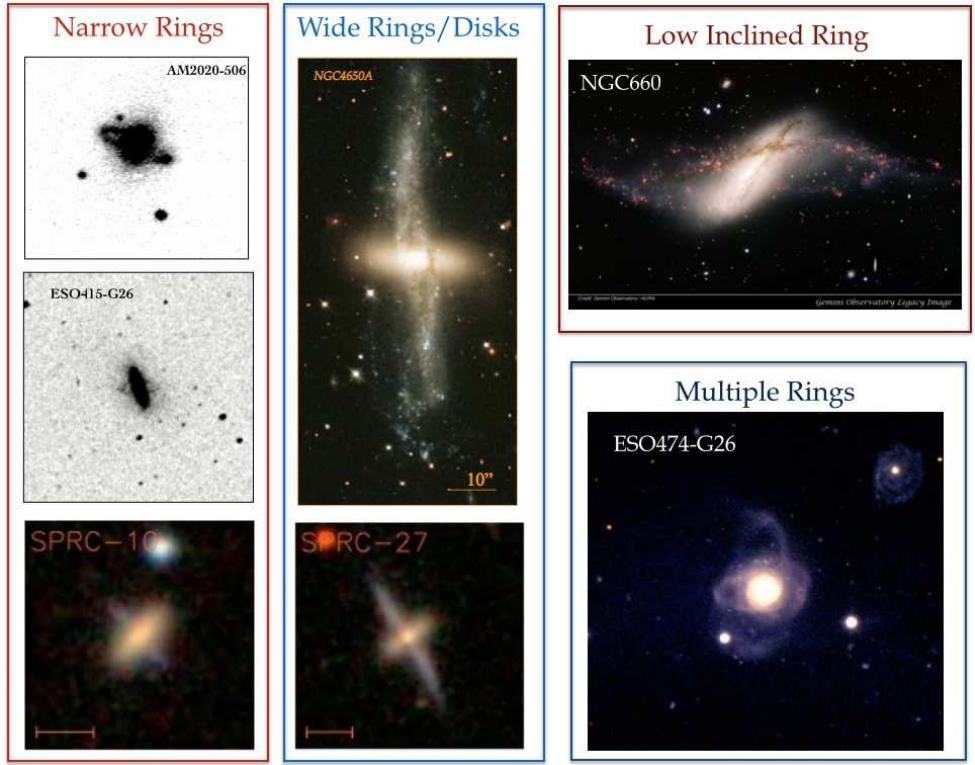}
\caption{Different morphologies observed for PRGs. 
Left panels: Narrow PRGs. Central panels: Wide polar
rings/disks. Right panels: Low inclined ring (top panel) and multiple
rings (bottom panel).}
\label{fig1}
\end{figure*}

{\it Morphology and light distribution.\/} Observations of PRGs show
that the morphology of the central HG resembles that of an early-type
galaxy (ETG). The polar ring is made up of gas, stars and dust that
orbits in a nearly perpendicular plane with respect to central HG.
While for most of PRGs the morphology of the HG is always similar to a
spheroidal galaxy, the polar structure has different shapes,
inclinations and extensions \citep{Whitmore1990, Moiseev2011}. In {\it
narrow\/} PRGs the polar ring has a smaller radius with respect to the
semi-major axis radius of the HG \citep[see e.g., ESO~415-G26 and
AM~2020-504;][]{Iodice2002a}. On the other hand, {\it wide\/} polar
structures are much more extended than the central
component \citep[like A~0136-0801, UGC~7576, and
NGC~4650A;][]{Iodice02, Iodice2002a, Cox2006}. In {\it multiple
ring\/} galaxies several decoupled ring-like structures are observed,
and at least one of them is on the polar direction with respect to the
central host \citep[e.g., ESO~474-G26;][]{Resh2005,Spavone2012}.  {\it
Low-inclined rings\/} are also related objects to the class of PRGs,
even if the angle between the HG and ring is less than 90\deg,
reaching also 45\deg\ as in NGC~660 \citep{vanDriel1995}. Some of the
PRG types cited above are shown in Fig.~\ref{fig1}. In both narrow and
wide PRGs, the surface brightness profiles along the major axis of the
HG are well reproduced by a S\'ersic law, with an $n$ exponent that
varies in the range $2 \le n \le 4$. On the contrary, light
distribution along the polar structure is quite different in narrow
rings with respect to the wide rings/disks: in the first case, the
ring appears as a `peak' on the underlying HG surface brightness,
while wide polar rings/disks have an extended exponential-like
decreasing profile \citep{Resh1994, Iodice02, Iodice2002a,
Iodice2002b, Spavone2012}.

\begin{figure*}[!t]
\centering
\includegraphics[width=12cm,keepaspectratio]{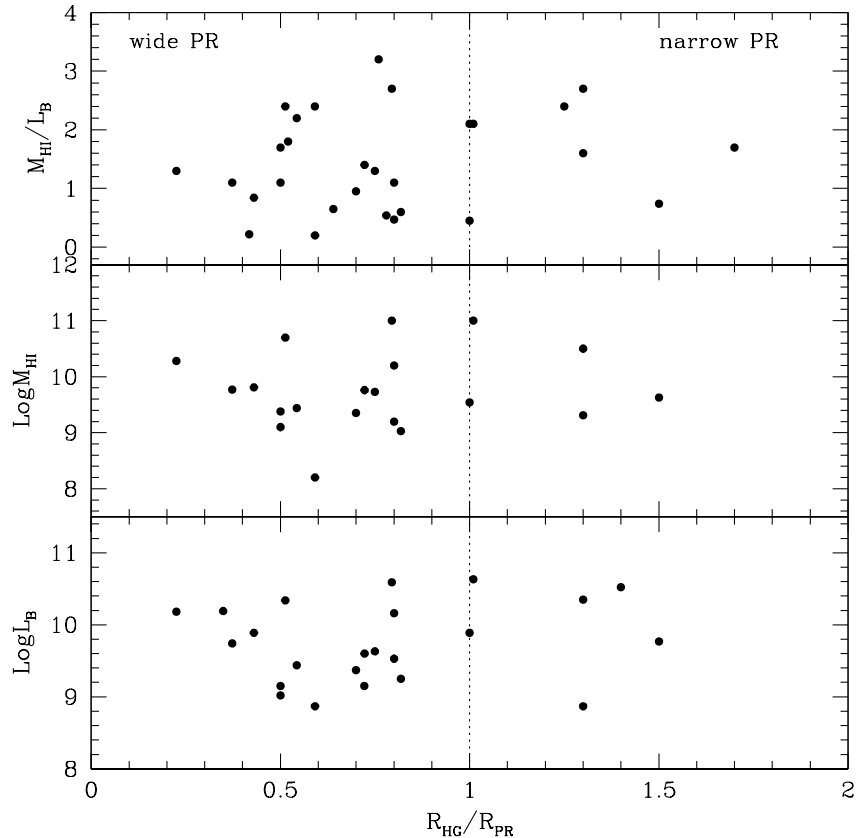}
\caption{Total luminosity $L_B$ (bottom panel),  
 \ion{H}{I} mass $M_{\rm HI}$ (middle panel) and ratio $M_{\rm
 HI}/L_B$ (top panel) as function of the relative radial extension
 $R_{\rm HG}/R_{\rm PR}$ of HG to polar ring. Data are
 from \citet{vanDriel2000, vanDriel2002}}
\label{fig2} 
\end{figure*}

{\it Total luminosity and H\,I distribution.\/} Fig.~\ref{fig2} shows
the distribution of the total luminosity $L_B$, \ion{H}{I} mass
$M_{\rm HI}$ and ratio $M_{\rm HI}/L_B$ as function of the relative
radial extension of the HG to the polar ring, $R_{\rm HG}/R_{\rm PR}$,
for a sample of PRGs \citep{vanDriel2000, vanDriel2002}.  In this
sample, wide PRGs are more numerous than narrow PRGs: this is
reasonably due to a selection effect, since wide polar structure in
galaxies are easily detectable, and more stable, than the narrow ones.
The distribution of all three quantities does not differ between
narrow and wide PRGs. For both types of galaxies, $L_B$ and $M_{\rm
HI}$ are uniformly distributed and vary in the ranges $8.8 \times
10^{10} \le L_B \le 10.8 \times 10^{10}$~L$_\odot$ and $8 \times
10^{10} \le M_{\rm HI} \le 11 \times 10^{10}$~M$_\odot$. Both narrow
and wide PRGs have on average a large amount of \ion{H}{I}, that can
be 2 or 3 times the total luminosity.
 
{\it Stellar kinematics.\/} As will be discussed in the next two
sections, the kinematics of both components (HG and ring/disk) in a
PRG is a crucial parameter to discriminate among different formation
scenarios for this class of objects, as well as for any morphological
type of galaxies. Unfortunately, there are few available spectroscopic
data for PRGs that avoid to perform a statistically significant
analysis. Anyway, I discuss two different cases: the narrow PRG
AM~2020-504 (Fig.~\ref{fig1}, left panel) and the wide polar disk
galaxy NGC~4650A (Fig.~\ref{fig1}, central panel), where the
kinematics along the major axis of the HG and along the polar
structure was published.  The rotation curve along the ring major axis
in AM~2020-504 is consistent with a ring, probably warped towards
outer radii \citep{Arnaboldi1993,Freitas}. In NGC~4650A, the rotation
curve is more similar to that of disks, with a rapid increment in the
inner regions followed by a flatter profile at larger
radii \citep{Swaters2003}. The different kinematics observed for the
narrow polar ring and wide polar disk turns to be consistent with the
different morphology and light distribution.  This is not the case for
the central component: even if, in both galaxies, the HG has a
spheroidal shape and similar light profile, the observed kinematics is
quite different. For both galaxies, along the major axis of the HG
rotation velocity increases and reaches $v\simeq100-130$~km~s$^{-1}$
in AM~2020-504, and $v\simeq80-110$~km~s$^{-1}$ in
NGC~4650A \citep{Arnaboldi1993, Iodice2006}. On the other hand, the
velocity dispersion profile for AM~2020-504 is very similar to those
typically observed in ETGs, peaked in the center ($\sigma \simeq
260$~km~s$^{-1}$) and decreasing outwards \citep[$\sigma \simeq
100$~km~s$^{-1}$,][]{Arnaboldi1993}. For NGC~4650A, the
high-resolution Very Large Telescope (VLT)
spectroscopy \citep{Iodice2006} have revealed that the velocity
dispersion remains almost constant at $\sigma \simeq 60$~km~s$^{-1}$
out to the last measured data points.

{\it Colors, ages, and baryonic mass.\/} Both narrow and wide polar
structures are on average bluer and younger than the central
HG \citep{Resh1994, Iodice02, Iodice2002b}, which is free of dust and
gas. The large amount of neutral and ionized gas reside in the polar
component \citep[e.g.,][]{Cox2006}. Colours, ages, and gas content in
the HG and polar structure of AM~2020-504 and NGC~4650A are compared
in Tab.~\ref{tab1}.  The $B-K$ colour, age estimate, and baryonic mass
(i.e., mass of stars plus gas) are comparable for the HG in both
galaxies. On the other hand, the narrow ring in AM~2020-504 is redder
and older than the wide polar disk in NGC~4650A. Furthermore, the
baryonic mass in NGC~4650A is 2 times larger than that in AM~2020-504.

\begin{table}[t!]
\caption{Colors, ages, and baryonic mass for the narrow 
PRG AM~2020-504 and wide polar disk galaxy NGC~4650A.}
\smallskip
\begin{center}
{\small
\begin{tabular}{cccc}
\tableline
\noalign{\smallskip}
PRG & $B-K$ & Age & $M_{\mbox{star+gas}}$ \\
         & [mag] & [Gyr] & [$10^9$~M$_{\odot}$] \\
\noalign{\smallskip}
\tableline
\noalign{\smallskip}
 & Host Galaxy & & \\
\noalign{\smallskip}
\tableline
\noalign{\smallskip}
NGC~4650A & 2.86 & $1-3$ & 5 \\
AM~2020-504 & 3.5 & $3-5$ & 6 \\
\noalign{\smallskip}
\tableline
\noalign{\smallskip}
 & Polar Structure & & \\
\tableline
\noalign{\smallskip}
NGC~4650A & 1.6 & 0.1 & 12 \\
AM~2020-504 & 2.7 & 1 & 5 \\
\noalign{\smallskip}
\tableline
\end{tabular}
}
\end{center}
\label{tab1}
\end{table}

\begin{figure*}[!t]
\centering
\includegraphics[width=11cm,keepaspectratio]{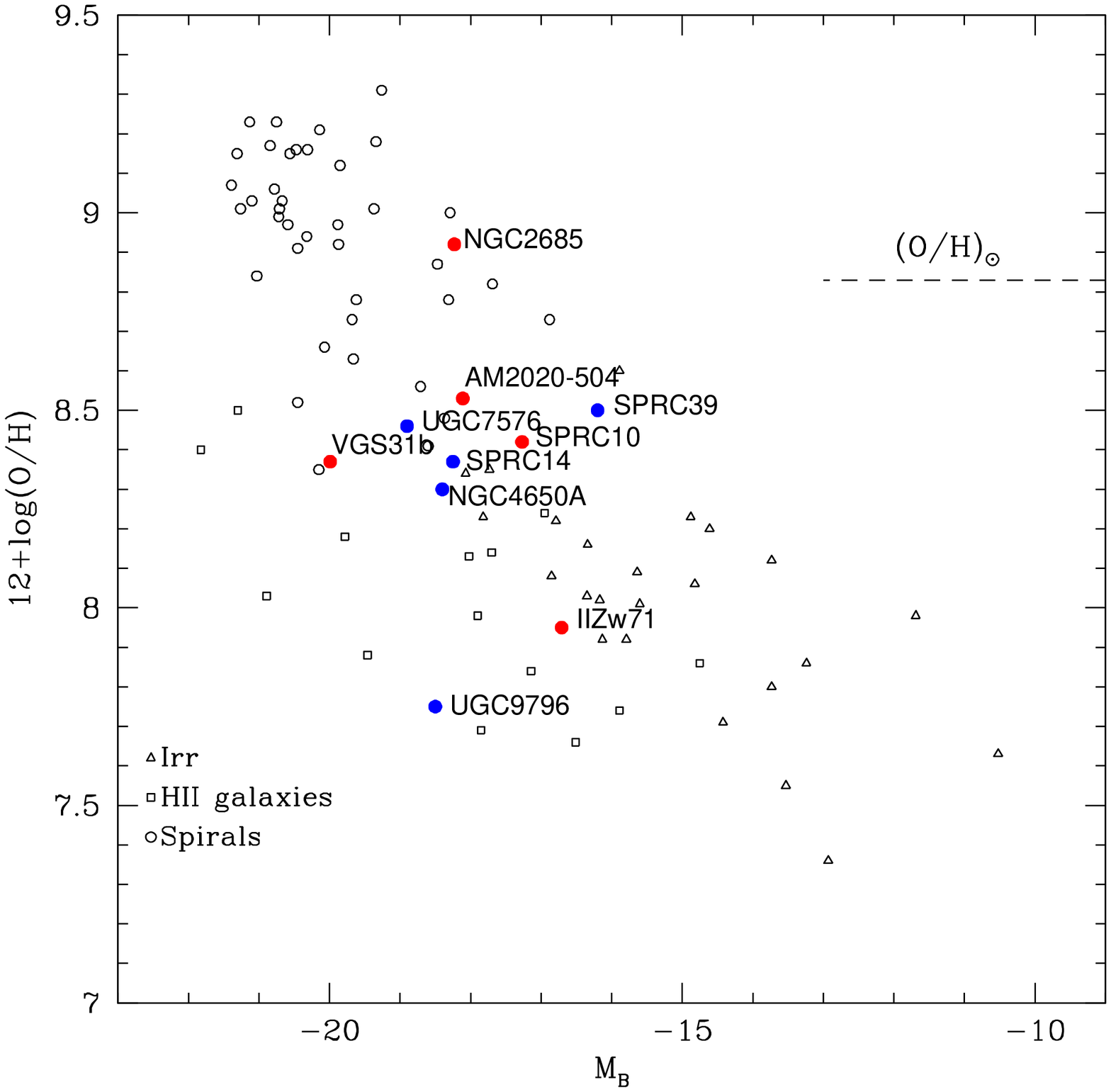}
\caption{The value of $12+\log({\rm O/H})$ as a function of the 
total $B$ luminosity for a sample of PRGs, including the wide polar
disk NGC~4650A \citep{Spavone2010} and the narrow PRG
AM~2020-504 \citep{Freitas}. The red and blue filled points correspond
to narrow and wide PRGs, respectively \citep[data are from][and
Moiseev et al., this volume]{Per09, EP1997, Spavone2010, Spavone2011,
Freitas, SI2013}. The sample of late-type disk galaxies
(spiral, \ion{H}{II}, and irregular galaxies) are by \citet{KZ1999}.}
\label{fig3} 
\end{figure*}

{\it Metallicity in the polar structure.\/} Recently, the new ongoing
field of research on PRGs aims to study the chemical abundances in the
polar structure. As I will discuss in the next section, this is
another fundamental physical parameter to discriminate among different
formation scenarios for this class of objects.  The emission lines
([\ion{O}{II}], H$\beta$, [\ion{O}{III}], H$\alpha$) were used to
derived oxygen abundance, expressed as $12+\log({\rm O/H})$,
metallicity $Z$, and global star formation rate for the polar
structure in several PRGs \citep[][and Moiseev et al., this
volume]{EP1997, Per09, Spavone2010, Spavone2012, Freitas, SI2013}. In
Fig.~\ref{fig3} is shown $12+\log({\rm O/H})$ vs. total $B$ luminosity
estimated for a sample of PRGs, compared with those derived for
late-type and disk galaxies. Most of PRGs with narrow and wide polar
structures have on average a sub-solar metallicity with $8.2 \le
12+\log({\rm O/H}) \le 8.6$ dex, similar to low-$Z$ spirals and
high-$Z$ irregular galaxies. There are three PRGs out of this range:
two of them have a very low oxygen abundance, $12+\log({\rm O/H}) < 8$
dex, the wide PRG UGC~9796 and the narrow PRG
IIZw71 \citep{Spavone2012, Per09}, and the spindle galaxy NGC~2685
with a metallicity comparable with the typical values for spiral
galaxies \citep{EP1997}.  It seems that there is no significant
difference in the metallicity between the narrow polar rings and wide
polar rings/disks. Anyway, the sample of PRGs with available measures
for oxygen abundances contains too few objects to give a definitive
conclusion on this subject, and more data need to be collected.

{\it Tully-Fisher relation and Faber-Jackson for PRGs.\/} As I
reviewed before, the polar structure in a PRG has very similar
properties (e.g., colors, \ion{H}{I} content, age, and metallicity) to
late-type galaxies, and in the case of the wide polar rings/disks
(like NGC~4650A) also the morphology and kinematics. On the other
hand, the observed morphology, colors and light distribution of the HG
in PRGs are very similar to that of an ETG (elliptical or S0), and, in
same cases (like AM~2020-504), also the kinematics.

How PRGs compare with the most important scaling relations for disks
and sphe\-ro-ids, i.e., the Tully-Fisher \citep[TF;][]{TF} and
Faber-Jackson \citep[FJ;][]{FJ} relations?  The position of the PRGs
with respect to the TF relation for bright disks was investigated
by \citet{Iodice2003} and, recently, for a larger sample
by \citet{Combes2013}. Most PRGs lie on a parallel relation with
respect to the TF for spirals, showing larger \ion{H}{I} linewidths
than expected for the observed total luminosity (Fig.~\ref{fig4}, left
panel). Numerical simulations suggest that this observational evidence
is related to the shape and orientation of the dark halo (DH) in these
systems: the larger rotation velocities observed in PRGs can be
explained by a flattened DH, aligned with the polar structure (see
also Combes, this volume). Furthermore, the new interesting result is
that in the $\log{\Delta V}-M_B$ plane the narrow and wide PRGs share
the same relation, without any segregation effect between the two
morphological types.

On the right panel of Fig.~\ref{fig4}, PRGs are compared with a sample
of ETGs in the plane $\log{\sigma}-M_B$, where $\sigma$ is the central
velocity dispersion of the HG. Differently from what observed for the
TF, both narrow and wide PRGs are far from the FJ relation for ETGs,
furthermore, it seems that exist a bimodal distribution of PRGs in the
plane $\log(\sigma)-M_B$, with some HG having higher $\sigma$ and few
others lower $\sigma$ with respect to spheroids of comparable total
luminosity.  If this result will be confirmed by using a larger sample
of PRGs, it is a relevant observational fact for the formation
scenarios.

\section{Formation Mechanisms for Polar Ring Galaxies}\
\label{theory}

Up to date three main formation scenarios have been proposed for PRGs
(see also Combes, this volume). A major dissipative merger, where a
PRG results from a `polar' merger of two disk galaxies with unequal
mass \citep{Bek98, Bou05}. The tidal accretion of material (gas and/or
stars) by outside, where a polar ring/disk may form by 1) the
disruption of a dwarf companion galaxy orbiting around an early-type
system, or by 2) the tidal accretion of gas stripped from a disk
galaxy outskirts, captured by an ETG on a parabolic
encounter \citep{ReshSot1997,Bou03,Han09}.  Recently, in the framework
of disk formation, the cold accretion of pristine gas along a filament
has been proposed as possible formation process for a wide polar disk,
like NGC~4650A \citep{Mac06, Brook08}. \citet{Snaith2012} have
revisited the cold accretion scenario, giving more stringent
constraints on the structure of the polar disk: this component should
have a sub-solar metallicity $Z= 0.2$~Z$_{\odot}$, a flat metallicity
gradient, and a younger age ($<1$~Gyr) with respect to that of the HG
($4-5$~Gyr). Furthermore, the DH has its major axis aligned along the
polar disk, it is rounder in the inner regions ($c/a=0.93$) and has an
increasing flattening towards larger radii ($c/a=0.67$).

How one can discriminate among different formation processes?  As for
any galaxy formation mechanism, the proposed scenarios need to account
for the morphology, gas content, star and gas kinematics, and
stability. Therefore, the key physical parameters that let to
discriminate among the three formation scenarios are 1) the baryonic
mass (stars plus gas) ratio between HG and polar structure; 2) the
star motions in the HG, i.e., rotation velocity and velocity
dispersion; 3) the metallicity and star formation rate in the polar
structure.

\section{Observations versus Theoretical Predictions}
\label{conf}

The wide polar disk galaxy NGC~4650A is the best studied PRG, since a
large wealth of data are available for this galaxy (see
Sect.~\ref{obs}), including \ion{H}{I} data \citep{Arnaboldi1997},
near-infrared and optical photometry \citep{Iodice02}, and long slit
kinematics \citep{Swaters2003, Iodice2006}. Thus, for this object all
the key physical parameters were derived and it can be considered the
first test case to discriminate among different formation scenarios
for PRGs.  NGC~4650A has a large baryonic mass in the polar structure,
which is equal or even larger than that in the HG \citep[][see also
Tab.~\ref{tab1}]{Iodice02}. The central spheroid is a
rotationally-supported system with the maximum rotation velocity
$v_{\rm{rot}} \simeq 80 - 100$~km~s$^{-1}$ \citep{Iodice2006}. The
polar disk has a sub-solar metallicity $Z = 0.2$ Z$_{\odot}$ and there
is no metallicity gradient along this component \citep[][see also
Fig.~\ref{fig3}]{Spavone2010}. By comparing the above observed
quantities with those expected by each formation mechanism, the
following conclusions were derived for NGC~4650A: 1) the merging
scenario is ruled out because, according to
simulations \citep[e.g.,][]{Bou05}, it fails to form massive polar
disk around an HG with rotation velocities as large as observed along
the HG major axis; 2) both the large baryonic mass in the polar
structure and its large extension can not be reconciled with a polar
disk formed via the gradual disruption of a dwarf satellite galaxy; 3)
the tidal accretion could form a such massive and extended polar disk
if the donor galaxy has a large amount of gas at large radii (where it
is not strongly gravitationally bound) and outside the stellar
disk \citep[e.g.,][]{Bou03}; 4) the measured metallicity and its flat
gradient, together with the polar disk and HG morphologies,
kinematics, gas content, and stellar ages, turn to be consistent with
the predictions by \citet{Snaith2012} for a polar disk formed through
the accretion of external cold gas from cosmic web
filaments \citep{Spavone2010}.

\begin{figure*}[!t]
\centering
\includegraphics[width=6.5cm,keepaspectratio]{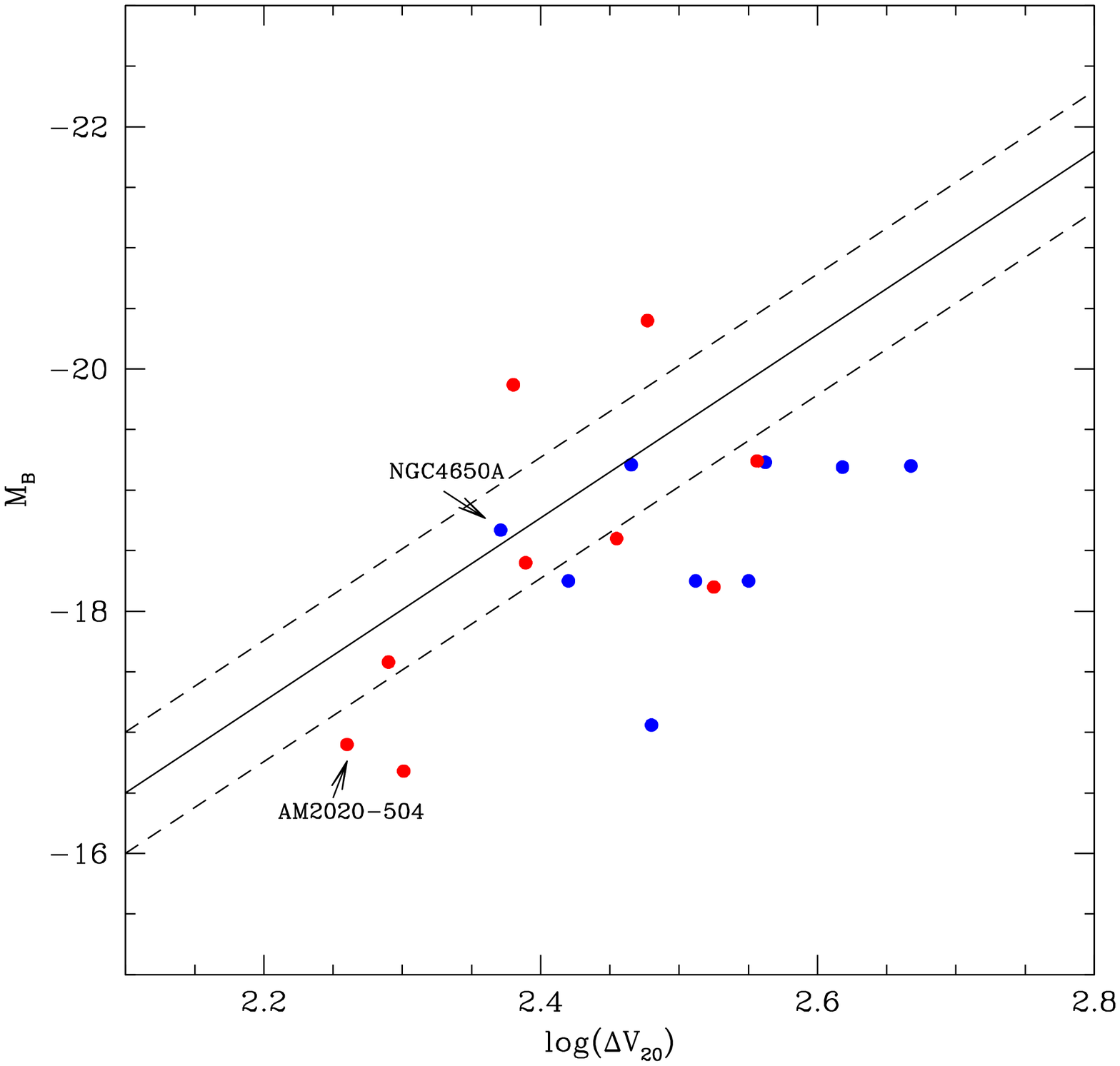}
\includegraphics[width=6.5cm,keepaspectratio]{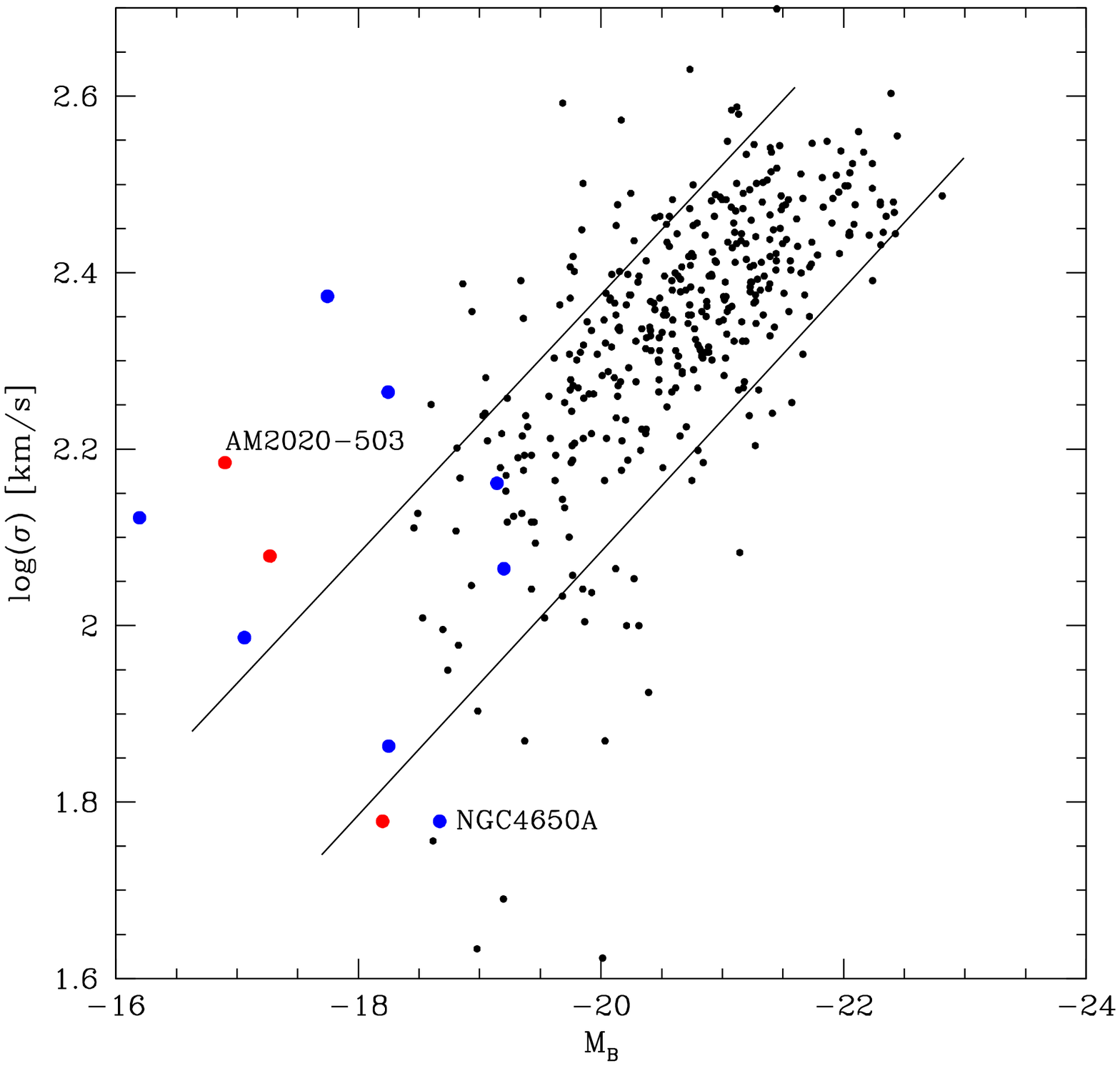}
\caption{TF and FJ relations for narrow (red circles) and wide 
(blue circles) PRGs. Letft panel: Absolute magnitude in the $B$ band
as a function of the linewidth at $20\%$ of the peak \ion{H}{I}-line
flux density ($\Delta V_{20}$), for a sample of PRGs \citep[data are
from][]{Iodice2003, Combes2013}, compared with the average TF relation
for a sample of spirals from \citet{Giovanelli1997}. The solid line is
the linear interpolation of the TF relation for spirals and the dashed
lines indicate the limit where the $81\%$ disks lie inside. Right
panel: Absolute magnitude in the $B$ band as a function of the central
velocity dispersion $\sigma$ of the HG in a sample of PRGs \citep[data
are from][and Moiseev, priv.  comm.]{Whitmore1990, Iodice2006}. Black
points are for the sample of ETGs by \citet{Focardi2012}. }
\label{fig4} 
\end{figure*}

By analyzing the available data for the narrow PRG AM~2020-504 (see
Sect.~\ref{obs}), the same analysis done for NGC~4650A can be
performed. Observations show that the HG and ring have comparable
baryonic mass (Tab.~\ref{tab1}); the HG has a high central velocity
dispersion along the major axis that decreases outwards, while the
rotation velocity increases out to $\sim100-130$~km~s$^{-1}$; the ring
has similar oxygen abundances to those of low-$Z$ spiral
galaxies \citep[][see also Fig.~\ref{fig3}]{Freitas}. Thus, taking
into account the above observed properties, and the HG morphology,
colors, and age which are very similar to those of an ETG, the ring in
AM~2020-504 could be reasonably formed through tidal accretion of
material from outside by a pre-existing ETG. This process could not
modify the global structure of the progenitor accreting galaxy, which
remains with a spheroidal morphology and same colors and age. On the
contrary, as suggested by the position of AM~2020-504 with respect to
the FJ for ETGs (Fig.~\ref{fig4}), the kinematics could change: the HG
has an higher velocity dispersion of stars with respect to the typical
values observed for ETGs of comparable total luminosity. Simulations
of galaxy formation show that gravitational interactions and merging
affect the observed kinematics of the remnant
galaxy \citep[e.g.,][]{Bou05,Naab2006}. This could be also a
reasonable explanation for the smaller $\sigma$ observed in the HG of
NGC~4650A in the $\log\sigma-M_B$ plane (Fig.~\ref{fig4}). If the HG
was originally a disk \citep[as assumed by the cold accretion scenario
proposed by][]{Brook08} and thus was characterized by a very low
velocity dispersion, the gravitational interaction that let to the
formation of the polar disk could have `puffed-up' the disk, which now
appears in its final stage as a spheroidal system with a larger
$\sigma$.

A similar kind of analysis has been performed for other PRGs to test
the accretion scenarios versus the merging
process \citep[e.g.,][]{Resh2005, Spavone2011, Spavone2012, SI2013}.

\section{Concluding Remarks}

In this paper I reviewed the latest results obtained for PRGs from
both observational and theoretical studies. This class of galaxies
includes different kind of morphologies, narrow rings, wide polar
rings/disks, and related objects (e.g., multiple or low inclined
rings). I focused on the analysis of the observed physical properties
(e.g., structure, colours, age, metallicity, and kinematics) for
narrow and wide PRGs. In particular, I compared AM~2020-504 and
NGC~4650A, which are the two prototypes for narrow and wide PRGs,
respectively.  I discussed similarities and differences between the
two kinds of systems and how they reconcile with the main formation
scenarios proposed for PRGs. The main results of this analysis are
listed below.

\begin{itemize}

\item For both narrow and wide PRGs, total luminosity $L_B$ 
and \ion{H}{I} mass $M_{\rm HI}$ are uniformly distributed and vary in
the ranges $8.8 \times 10^{10} \le L_B \le 10.8 \times
10^{10}$~L$_\odot$ and $8 \times 10^{10} \le M_{\rm HI} \le 11 \times
10^{10}$~M$_\odot$. Both narrow and wide PRGs have, on average, a
large amount of \ion{H}{I}, that can be 2 or 3 times the total
luminosity.

\item The HG has similar (spheroidal) morphology, colours, and age in both
AM~2020-504 and NGC~4650A, but very different kinematics. In
particular, the profile of the stellar velocity dispersion is
different. As a consequence, the two galaxies have a different
position with respect to the FJ relation for ETGs (see
Fig.~\ref{fig4}, right panel). This is also observed for other few
PRGs which seem to have a bimodal distribution in the
$\log{\sigma}-M_B$ plane, with some HGs having a larger $\sigma$ and
few others a lower $\sigma$ with respect to spheroids of comparable
total luminosity. This bimodality seems to be independent from the
morphological type (narrow or wide PRGs). This result needs to be
confirmed with a larger sample of PRGs, since its implications on the
formation scenarios for spheroids are quite relevant.

\item Contrary to the HGs, the polar structures 
in narrow and wide PRGs have different morphology, baryonic mass, and
kinematics, but, on average, similar oxygen abundance
(Fig.~\ref{fig3}). NGC~4650A and AM~2020-504 slso show different age
and colours, being the narrow ring older and redder than the wide
polar disk (see Tab.~\ref{tab1}). Furthermore, the position of both
types of PRGs with respect to the TF relation for spiral galaxies is
on average the same, i.e., most PRGs lie on a parallel relation with
respect to that for spirals, showing larger \ion{H}{I} linewidths than
expected for the observed total luminosity (see Fig.~\ref{fig4}, left
panel).

\item The key physical parameters (i.e., baryonic mass, HG kinematics,
and metallicity in the polar structure), that let to discriminate
among the three formation scenarios suggested for PRGs (see
Sect.~\ref{theory}) were derived in several works for some PRGs,
including NGC~4650A and AM~2020-504. By comparing them with the
predictions from models, one can trace the formation history for each
object. In particular, in the case of NGC~4650A the cold accretion of
gas along a filament is successfully tested.

\end{itemize}

All the above results suggest that there is not a unique formation
scenario for PRGs, but different mechanisms need to be invoked to
explain the whole range of the observed properties for this class of
galaxies.  As already showed by \citet{Bou03}, the analysis performed
in this review confirms that both major merger and tidal accretion
scenarios are able to account for the observed morphologies and
kinematics of narrow as well as wide polar rings/disks.  The
simulations of a cold accretion of pristine gas along a filament
account for the formation of a polar disk, like NGC~4650A. In the
metallicity - luminosity plane (see Fig.~\ref{fig3}), there are other
PRGs, both narrow and wide polar rings/disks, with similar or even
lower oxygen abundance: if the low metallicity is a strong constraint
in favour of the cold accretion, may polar rings (not disks!), form
through this mechanisms?

\acknowledgements 
I wish to thank A. Moiseev for providing the unpublished kinematic
data for some PRGs of the SPRC catalogue, which turn to be very useful
to produce the FJ plot presented in this paper. I would like to warmly
thank M.~Arnaboldi, D.~Bettoni, F.~Bournaud, M.~Capaccioli,
L.~Coccato, F.~Combes, E.~M.~Corsini, T.~de Zeeuw, K.~C.~Freeman,
J.~S.~Gallagher, G.~Galletta, O.~Gerhard, P.~Kroupa, A.~Moiseev,
N.~R.~Napolitano, R.~Saglia, L.~S.~Sparke, and M.~Spavone for the
discussions and/or works made together in the last ten years of my
career, that gave a huge contribution to my knowledge on the research
field presented in this work.


\end{document}